\documentclass[10pt]{iopart}
\lccode`\3`\3
\lccode`\/`\/
\usepackage[
    bottom=1.1cm
]{geometry}

\expandafter\let\csname equation*\endcsname\relax
\expandafter\let\csname endequation*\endcsname\relax
\usepackage{amsmath}
\usepackage{amssymb}
\usepackage{dblfloatfix}
\usepackage[per-mode = symbol]{siunitx}
\usepackage{enumitem}
\usepackage{graphicx}
\usepackage{xcolor}
\usepackage{siunitx}
\usepackage{bm}
\usepackage[
    colorlinks,
    citecolor=blue,
    urlcolor=black,
    linkcolor=blue
]{hyperref}
\usepackage{droidsansmono}
\usepackage{subcaption}
\DeclareCaptionFormat{iop-caption}{\raggedright \footnotesize \textbf{#1#2}#3}
\DeclareCaptionFormat{iop-subcaption}{\centering\footnotesize #1#2#3}
\usepackage{listings}

\usepackage{tudacolors}
\usepackage{pgfplots}
\usepackage{pgfplotstable}
\usepgfplotslibrary{colormaps} 
\usepackage{calc}

\usepackage{siunitx}
\usepackage{multirow}

\usepackage{macros}

\makeatletter
\def\thebibliography#1{\list
 {\hfil[\arabic{enumi}]}{\topsep=0\p@\parsep=0\p@
 \partopsep=0\p@\itemsep=0\p@
 \labelsep=5\p@\itemindent=0\p@
 \settowidth\labelwidth{\footnotesize[#1]}%
 \leftmargin\labelwidth
 \advance\leftmargin\labelsep
 \usecounter{enumi}}\footnotesize
 \def\newblock{\ }
 \sloppy\clubpenalty4000\widowpenalty4000
 \sfcode`\.=1000\relax}
\makeatother

\begin{document}

\title{Homogenization of {HTS} coils with the h, h-phi, and t-omega foil conductor model}

\author{Elias Paakkunainen\textsuperscript{1,2,*}, Louis Denis\textsuperscript{3}, Benoît Vanderheyden\textsuperscript{3}, Christophe Geuzaine\textsuperscript{3}, Paavo Rasilo\textsuperscript{2}, and Sebastian Schöps\textsuperscript{1}}

\address{\textsuperscript{1}Institute for Accelerator Science and Electromagnetic Fields, Technical University of Darmstadt, Darmstadt, Germany}
\address{\textsuperscript{2}Electrical Engineering Unit, Tampere University, Tampere, Finland}
\address{\textsuperscript{3}Department of Electrical Engineering and Computer Science, Montefiore Institute, University of Liège, Liège, Belgium}
\ead{elias.paakkunainen@tu-darmstadt.de}
\hspace{73pt}{\footnotesize \textsuperscript{*}Corresponding author.}
\vspace{10pt}

\begin{abstract}
Efficient numerical models are required for the design of systems with high temperature superconductor (HTS) coils, as fully resolved finite element simulations of individual coated conductors become computationally prohibitive. This work applies the foil conductor model (FCM) to insulated HTS coils using magnetic field conforming $h$-(full), $h$-$\phi$, and $t$-$\omega$ formulations. The approach replaces individual turns by a homogenized bulk and ensures physically consistent current density distributions in the coils by using additional voltage basis functions in the finite element formulations. The models are verified in 2D axisymmetric and 3D geometries with a pancake coil simulation under AC transport current excitation. All FCM formulations show excellent agreement with reference detailed simulations, with coefficients of determination above 0.99 for instantaneous AC losses. In 3D, the $h$-$\phi$ and especially the $t$-$\omega$ formulation substantially reduce the number of degrees of freedom by using the magnetic scalar potential in non-conducting regions. Scalability is demonstrated with a 3D stack of racetrack coils model with a field- and angle-dependent critical current density. For the stack of racetrack coils, while maintaining accurate loss prediction, the $t$-$\omega$ FCM achieves a speedup factor of 22 and reduces degrees of freedom by \SI{78}{\percent} with respect to a detailed reference model.
\end{abstract}

%
\vspace{2pc}

\par\vspace{1.5pc}
\begingroup
\setlength{\fboxsep}{6pt}
\noindent\fbox{%
    \parbox{\dimexpr\textwidth-2\fboxsep-2\fboxrule\relax}{%
        \footnotesize
        This work has been submitted to a journal for possible publication. Copyright may be
        transferred without notice, after which this version may no longer be accessible.%
    }%
}
\par\endgroup

%
\maketitle
%
\ioptwocol

\section{Introduction}\label{sec:introduction}
High temperature superconductors (HTS) are intensively studied for multiple applications \cite{Coombs_2024aa}. These include, e.g., magnets for particle accelerators \cite{Bottura_2022ab} and fusion reactors \cite{Hartwig_2024aa}, electrical machines \cite{Haran_2017aa}, fault current limiters \cite{Sotelo_2022aa} and power cables \cite{Noe_2026aa}. The design of these applications benefits from efficient and accurate numerical models to reduce expensive and time-consuming experiments.

Often, the finite element method (FEM) is used for the numerical simulation of electromagnetic fields in HTS devices. Several formulations exist within the FEM framework along with separate numerical techniques. Comparisons of different approaches have been reported in the literature and we refer to, e.g., \cite{Dular_2020aa} and \cite{Dadhich_2024aa}. Notably, with the nonlinear material law of superconductors, expressing the material property in terms of the electrical resistivity instead of the conductivity has been observed to improve the numerical performance.

\begin{figure*}
    \centering
    \includegraphics[width=0.85\textwidth,trim=4.5em 0em 0em 0em,clip]{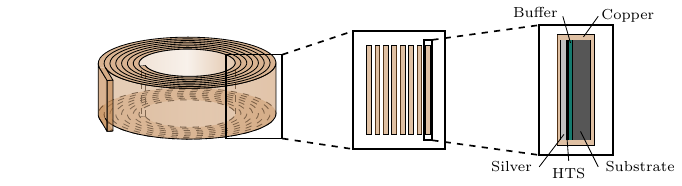}
    \caption{A sketch of the detailed geometry of a HTS pancake coil. The cross-section of the separate turns of the coil is shown alongside the layers of the CCs. Figure adapted from \cite{Paakkunainen_2024aa}.}
    \label{fig:pc_geom}
\end{figure*}

The large aspect ratio of HTS coated conductors (CCs) and the strong nonlinearity of the material properties make the simulations computationally expensive. Complex geometries of application-relevant designs and a large number of CCs in the HTS coils further increase the computational burden, particularly in 3D. Homogenization techniques are one possible approach to reduce the computational cost by neglecting the fine structure of the coils, i.e., the layers of the CCs and the individual coil turns. Figure~\ref{fig:pc_geom} illustrates the detailed geometry of a HTS pancake coil. In this work, the entire coil is modeled as a homogenized bulk material. In other works, homogenization has, e.g., been applied to the layers of the CCs in HTS coils while distinguishing individual turns~\cite{Schnaubelt_2023ab}. Also, the simultaneous multi-scale homogeneous models provide an alternative where some of the CCs are modeled in detail while the rest of the domain is homogenized \cite{Wang_2025ab,Denis_2026aa}. Lastly, reduced order methods can be used to describe the nonlinear material behaviour as demonstrated recently for systems with low \cite{Dular_2026aa} and high temperature superconductors \cite{Basei_2026aa}.

Homogenization of HTS coils has been studied with the $h$-(full) \cite{Zermeno_2013aa}, $t$-$a$ \cite{Berrospe-Juarez_2019aa} and $j$-$a$ formulations \cite{Wang_2023aa}, and more recently with the $j$-$a$-$\phi$ formulation \cite{Dos-Santos_2026aa}. Many works discuss 2D models but there are also examples of 3D models \cite{Zermeno_2014aa, Vargas-Llanos_2022aa}. These works examined insulated coils, which are also the focus of this paper. Homogenization of non-insulated HTS coils has also been examined in 2D \cite{Mataira_2020aa}.

Recently, the so-called foil conductor model (FCM) has been proposed for the simulation of insulated HTS coils \cite{Paakkunainen_2025aa}. This approach was originally introduced for the simulation of normal conducting foil windings \cite{De-Gersem_2001aa, Dular_2002aa}. The main difference between this approach and the method presented by Zerme\~{n}o et al. \cite{Zermeno_2013aa}, or the $t$-$a$ homogenized model, is in the way the current is imposed. As an alternative to setting current constraints for each of the subdomains in the coil region, a discretization of the gradient of the electric scalar potential can be introduced, as described in more detail in \cite{Paakkunainen_2025aa}. The original FCM was implemented for the $a$ formulation which was extended to the $j$-$a$ formulation to be more suitable for simulating HTS. In a recent contribution, the FCM was further extended to $h$-conforming formulations, i.e., the $h$, $h$-$\phi$, and $t$-$\omega$ formulations in normal conducting and 2D superconducting applications \cite{Denis_2025aa}.

In this paper, we extend the formulations presented in \cite{Denis_2025aa} to simulations of HTS coils, the main novelty being 3D simulations. In practice, 2D approximations are often not accurate for realistic applications, thus highlighting the need for simulation models applicable to general 3D geometries. In a 3D setting, the $h$-$\phi$ and $t$-$\omega$ formulations are particularly interesting due to their reduced number of degrees of freedom (DoFs) compared to the full $h$, $t$-$a$ and $j$-$a$ formulations. The rest of this paper is structured as follows. In section~\ref{sec:model}, the mathematical framework and finite element formulations are presented. In section~\ref{sec:results}, numerical results are shown for example simulations of a pancake coil and a stack of racetrack coils to verify the proposed models and to highlight the improvement in computational efficiency. Finally, conclusions are drawn in section~\ref{sec:conclusions}.

\section{Methods}\label{sec:model}
\begin{figure}
    \centering
    \includegraphics[width=\columnwidth]{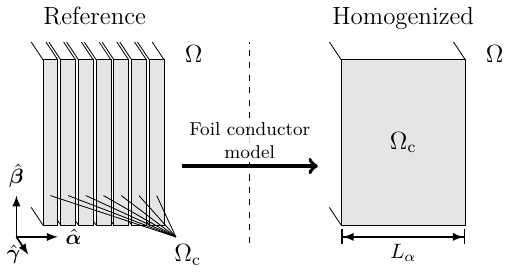}
    \caption{Principle of the foil conductor model, replacing distinct turns of the reference model with a single homogenized bulk, with the local coordinate system $\alpha$-$\beta$-$\gamma$ attached to the cross-section. Figure adapted from~\cite{Denis_2025aa}.}
    \label{fig:fcm-principle}
\end{figure}

This section recalls the conventional $h$-$\phi$ formulation~\cite{Dular_2020aa} used for reference simulations, before describing the application of $h$-conforming formulations for the FCM, proposed in~\cite{Denis_2025aa}, to HTS coils. We consider the simulation of $N_{\text{c}}$ stacked HTS CCs as shown in figure~\ref{fig:fcm-principle}, which are insulated and arranged in series, so that all turns carry the same current, denoted by $I_{\text{t}}$. Both the 2D reference and the FCM assume concentric turns, neglecting the intrinsic spiral formed by actual CCs wound in HTS coils. Together, the turns form the conducting subdomain $\Oc$ within the computational domain $\O$.

\subsection{Reference Model}
The reference model is based on the $h$-$\phi$ formulation detailed in~\cite{Dular_2020aa}. After finite element discretization, the weak formulation aims at finding the magnetic field $\bm{h}$ satisfying 
\begin{equation}
    \volInt{\mu\partial_t\h}{\h'}{\O} + \volInt{\rho\,\curl{\h}}{\curl{\h}'}{\Oc} = 0 \label{eq:href}
\end{equation}
for all test functions $\bm{h}'$, with $\mu$ the magnetic permeability and $\rho$ the electrical resistivity. In this work, in the absence of ferromagnetic materials, $\mu$ is equal to the permeability of free space $\mu_{0}$. The volume integral over $\O$ of the inner product of two fields is denoted by $\volInt{\cdot}{\cdot}{\O}$. Here, the gradient of the electric scalar potential $v$ is considered constant per turn: $\grad v = \sum_{i=1}^{N_{\text{c}}} V_i \grad v_{\text{s},i}$, with $\grad v_{\text{s},i}$ global basis functions (BFs)~\cite{Dular_1999aa}. The magnetic field is discretized with edge BFs in the conducting domain $\Oc$ and nodal BFs in the non-conducting domain $\Occ = \O \setminus \Oc$. Moreover, $N_{\text{c}}$ global edge cohomology BFs~\cite{Pellikka_2013aa}, commonly referred to as \textit{thick cuts}, are required to fix the net current~\cite{Dular_2020aa}.

The resistivity of coated conductors is modeled with the power-law~\cite{Rhyner_1993aa}:
\begin{equation}
    \rho_{\text{HTS}} = \frac{\ec}{\jc} \left( \frac{\norm{\vec{j}}}{\jc} \right)^{n-1},
    \label{eq:power_law}
\end{equation}
where $\j$ is the current density, $\jc$ the critical current density, $\ec$ the critical electric field commonly chosen to be \SI{e-4}{\volt\per\meter}, and $n$ the power-law exponent. Generally, $\jc$ and $n$ depend locally on the system state, e.g., the magnetic field and temperature. In this work, we assume $n$ to be constant and the different used $\jc$ dependencies are discussed in section~\ref{sec:results}.

\subsection{Foil Conductor Model}
\begin{table*}
    \caption{Comparison of the different $h$-conforming FCMs in terms of discretization, number $n_{\text{cuts}}$ of cuts to construct the function space, corresponding number $n_{\text{DoFs}}$ of DoFs with respect to $h$-(full), and coil resistivity.}
    \centering
    \renewcommand{\arraystretch}{1.1}
    \begin{tabular}{*{5}{c}}
    \br
    FCM & Discretization & $n_{\text{cuts}}$ & $n_{\text{DoFs}}$ & Coil resistivity \\
    \hline
    $h$-(full)
    & Edge BFs in $\O$ ($\rightarrow$ spurious $\rho_{\text{air}}$)
    & 0
    & -
    & Anisotropic \\
    $h$-$\phi$
    & Edge BFs in $\Oc$, nodal BFs in $\Occ$
    & 1 per coil
    & Lower
    & Anisotropic \\
    $t$-$\omega$
    & $\alpha$-Edge BFs in $\Oc$, nodal BFs in $\O$
    & 1 per coil + cut-like BFs
    & Lowest
    & Isotropic \\
    \br
    \end{tabular}
    \label{tab:comparison-sec2}
\end{table*}

The FCM replaces the distinct turns of HTS coils with a single homogenized bulk, as represented in figure~\ref{fig:fcm-principle}. To do so, the FCM approximates the gradient of the electric scalar potential with a continuous function~\cite{De-Gersem_2001aa, Dular_2002aa} along the CC stack thickness (the $\alpha$-coordinate in the local coordinate system):
\begin{equation}
    \grad v = \Phi(\alpha)\,\grad v_{\text{s}}, \label{eq:fcm-ansatz}
\end{equation}
with $\grad v_{\text{s}}$ a single global BF and $\Phi(\alpha)$ the \textit{voltage} distribution function. $\Phi(\alpha)$ is expressed in terms of its own BFs, which need not depend on the underlying mesh. The FCM $h$-conforming weak formulation~\cite{Denis_2025aa} reads:
find $\bm{h}$ and $v$, such that
\begin{multline}
    \volInt{\mu\partial_t\h}{\h'}{\O} + \volInt{\rho\,\curl{\h}}{\curl{\h}'}{\Oc} \\ + \volInt{\grad v}{\curl{\h}'}{\Oc} = 0 \label{eq:hfcm1}
\end{multline}
\begin{equation}
    \volInt{\curl \h}{\Phi' \grad v_{\text{s}}}{\Oc} - \frac{N_{\text{c}}}{L_{\alpha}}\int_{L_{\alpha}} \It\,\Phi'\,d\alpha = 0 \label{eq:hfcm2}
\end{equation}
holds for all test functions $\bm{h}'$, $\Phi'$, with $L_{\alpha}$ the stack thickness. While \eqref{eq:hfcm1} is a direct extension of the reference weak formulation~\eqref{eq:href}, the current per turn $\It$ is imposed weakly by \eqref{eq:hfcm2}.

Here, only the HTS layers are considered for the effective resistivity evaluation. In \eqref{eq:power_law}, $\jc$ is replaced by the engineering critical current density $\jceng = \lambda \jc$, where $\lambda$ is the fill factor of the HTS layer in the CC. Extension of the models to account for all the conducting layers of the CCs is possible by additionally solving the current sharing problem between the different layers in a similar way to what is presented in~\cite{Bortot_2020aa}.

Different $h$-conforming FCM formulations have been described in~\cite{Denis_2025aa}, namely the $h$-(full), the $h$-$\phi$ and the $t$-$\omega$ formulations. While they share the same weak formulation~\eqref{eq:hfcm1}-\eqref{eq:hfcm2}, they differ in the discretization of the magnetic field. Their comparison is summarized in table~\ref{tab:comparison-sec2}.

The $h$-(full) FCM, similarly to the commonly known $h$ formulation, is exclusively based on edge BFs. Therefore, it requires the definition of a spurious resistivity in the air~\cite{Dlotko_2019aa}, here set to $10^{-3}$~$\Omega$~m, with the air being then included in $\Oc$.

The $h$-$\phi$ FCM makes use of the magnetic scalar potential $\phi$ since $\bm{h} = - \grad \phi$ in the non-conducting domain. Its discretization is thus based on a combination of both edge and nodal BFs. Notably, it requires the introduction of a single cut per coil in contrast to reference models with concentric turns. The use of nodal BFs in $\Occ$ greatly reduces the size of the numerical system to be solved~\cite{Denis_2025aa}.

The $h$-(full) and $h$-$\phi$ FCM require the introduction of an anisotropic resistivity tensor in $\Oc$ to prevent spurious current sharing between neighbouring turns~\cite{Denis_2025aa}. In the local $\alpha$-$\beta$-$\gamma$ coordinates, it is expressed as
\begin{equation}
    \bm{\rho} = \begin{pmatrix}
        \rho_0 & 0 & 0 \\
        0 & \rho_{\text{HTS}} & 0 \\
        0 & 0 & \rho_{\text{HTS}}
    \end{pmatrix},
\end{equation}
with the spurious resistivity $\rho_0$ (ideally infinite, since the turns are insulated) set to $10^{-3}$~$\Omega$~m in this work.

In contrast, the $t$-$\omega$ FCM circumvents the anisotropic resistivity tensor by strongly enforcing zero current in the $\alpha$-direction perpendicular to the CCs. A dedicated discretization, detailed in~\cite{Denis_2025aa}, extends the magnetic scalar potential within the conducting domain, while only considering edge shape functions perpendicular to the CCs (along the $\alpha$-direction). This is similar to the definition of the electric vector potential $\bm{t}$ in the homogenized $t$-$a$ formulation~\cite{Vargas-Llanos_2022aa}. While the resistivity in~\eqref{eq:hfcm1} can then be evaluated as a scalar, this comes at the cost of introducing additional (yet very localized) cut-like BFs~\cite{Denis_2025aa}. As discussed in~\cite{Denis_2025aa}, this leads to a further reduction of the number of DoFs, and is thus expected to be more efficient than the $h$-(full) and $h$-$\phi$ formulations. All the presented FCM formulations use a structured hexahedral mesh in the homogenized coil domain.

\section{Numerical Results}\label{sec:results}
In this section, the models described in section \ref{sec:model} are implemented and verified for the test cases of a single pancake coil and a stack of racetrack coils. The geometries are generated and meshed with Gmsh~\cite{Geuzaine_2009ab}, and GetDP \cite{Dular_1998ac} is used as the FEM solver. All the implementations have been developed with open-source software and are made available in \cite{Paakkunainen_2026aa}. Across the different examined models, the parameters of the HTS CCs are kept the same, $\jc$ being the exception. The corresponding parameters are listed in table~\ref{tab:cc_params}. All the models are run in a compute node with Intel Xeon Platinum 8160 CPUs, and 32~MPI processes are used unless stated otherwise.
\begin{table}
    \caption{Parameters of the HTS CC.}
    \centering
    \renewcommand{\arraystretch}{1.1}
    \begin{tabular}{*{3}{c}}
    \br
    Quantity & Symbol & Value \\
    \hline
    Power law exponent & $n$ & \SI{25}{} \\
    CC thickness & - & \SI{100}{\micro\meter} \\
    CC width & - & \SI{12}{\milli\meter} \\
    Fill factor of HTS & $\lambda$ & \SI{0.01}{} \\
    \br
    \end{tabular}
    \label{tab:cc_params}
\end{table}

\subsection{Pancake Coil}

For verification purposes, a pancake coil with ${N_{\text{c}}=20}$ turns is considered both in 2D axisymmetric and 3D geometries. The inner radius of the coil is \SI{25}{\milli\meter}. The critical current density is assumed to be constant ${\jc=\SI{e10}{\ampere\per\meter\squared}}$. A sinusoidal current with an amplitude $\It=0.8\Ic$, proportional to the critical current $\Ic$, is imposed at \SI{50}{\hertz} frequency to the coil. The continuous voltage function $\Phi(\alpha)$ is approximated with a third-order global polynomial. Other choices for BFs are possible, also with local support, and have been discussed in \cite{Paakkunainen_2025aa}. With the 3D FCM, 1/8 of the coil geometry is simulated, whereas the 3D resolved model, which distinguishes all the individual turns, has to consider the whole geometry to account for the spiral shape of the wound CC (1/2 symmetry would be possible but is not utilized due to the availability of implementations). The extended symmetry properties are a benefit of the homogenization.

The 3D resolved model is based on the FiQuS Pancake3D module \cite{Atalay_2024aa}. The model approximates turn-to-turn contacts with perfectly insulating layers of vanishing thickness, consequently avoiding the volumetric meshing of the very thin layer and already leading to improved numerical performance. The insulating pancake coil is implemented with the perfectly insulating surface model \cite{Dular_2003ab} extension of the FiQuS Pancake3D module \cite{Wozniak_2025aa}. The Pancake3D module also includes the copper terminals as current leads, which are not included in the homogenized models so that symmetry planes can be utilized. With the 3D resolved model, the copper terminal resistivity is set artificially to a higher value of \SI{e-6}{\ohm\meter} to remove its effect on the field solution. Moreover, the current sharing computation~\cite{Bortot_2020aa} is deactivated in the 3D resolved model. The losses are computed only for the coil region. A large contact resistance between the copper terminals and the pancake coil ensures current enters the coil azimuthally in first and last turns.

\begin{figure}
    \centering
    \includegraphics[width=0.47\textwidth]{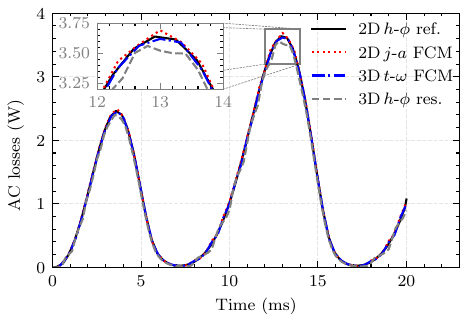}
    \caption{The instantaneous AC losses for the pancake coil models. The losses of the 2D axisymmetric models and 3D FCM are scaled for the entire coil geometry.}
    \label{fig:pc_losses}
\end{figure}

The different models are compared through the losses produced by the imposed AC current. The examined models are the 2D axisymmetric FCM with the $j$-$a$, $h$-(full), and $h$-$\phi$ formulations, and the 3D FCM with the $h$-(full), $h$-$\phi$, and $t$-$\omega$ formulations. The 2D reference and 3D resolved models use the $h$-$\phi$ formulation. Figure~\ref{fig:pc_losses} shows the instantaneous losses in the coil for some selected models.  The mesh was chosen to be fine enough so that both the 2D axisymmetric and the 3D models produce a smooth curve for the instantaneous AC losses. A good agreement of the models is observed, verifying the modelling approach. Also, for this example simulation, the assumption of concentric turns leads to a small difference with respect to the 3D resolved model with the accurate geometry.

\begin{figure*}
    \centering
    \includegraphics[width=0.8\textwidth]{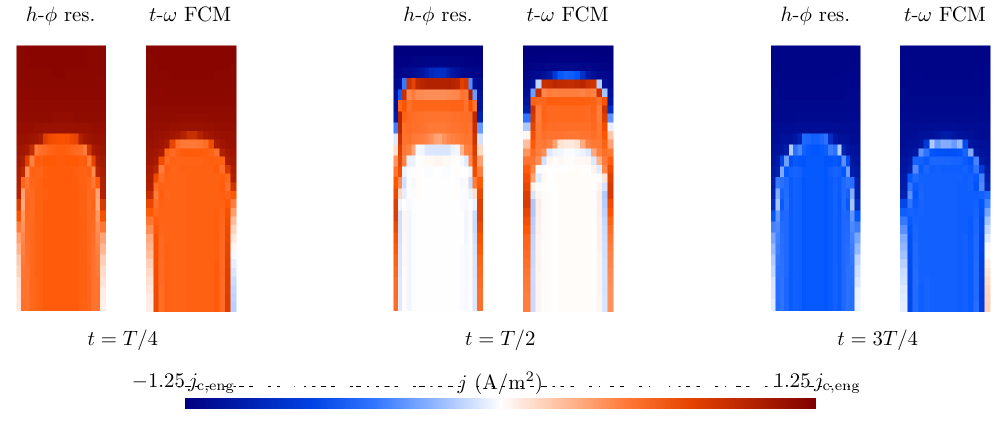}
    \caption{Current density distributions for a half cut plane of the pancake coil at different time instants. The current density is normalized with respect to the engineering critical current density $\jceng$. The distribution is shown for the 3D resolved model and the 3D $t$-$\omega$ FCM.}
    \label{fig:pc_j}
\end{figure*}

To quantify the accuracy, the coefficient of determination $R^{2}$ is computed for the instantaneous losses $p$ with respect to the 2D reference model as it is associated to the most refined mesh. The coefficient is calculated as 
\begin{equation}
    R^{2} = 1 - \frac{\displaystyle\int_{0}^{T}\left(p-p_{\text{ref}}\right)^{2}dt}{\displaystyle\int_{0}^{T}\left(p_{\text{ref}}-\bar{p}_{\text{ref}}\right)^{2}dt},
\end{equation}
where $p_{\text{ref}}$ is the instantaneous losses of the 2D reference model, $\bar{p}_{\text{ref}}$ its mean value and $T$ is the period of the imposed AC current. The computed $R^{2}$ values are listed in table~\ref{tab:pc_times}. The coefficients confirm the good agreement as the $R^{2}$ values are greater than 0.99 for all of the examined models. Figure~\ref{fig:pc_j} shows the distribution of the current density $\j$ through a half cut plane of the coil (half of the coil cross-section illustrated in figure~\ref{fig:pc_geom}) for the 3D resolved model and the 3D $t$-$\omega$ FCM at different time instants. A good agreement between the models is observed.

\begin{table*}
    \caption{Computation cost of the pancake coil models. 2D axisymmetric models are run with a single MPI process.}
    \centering
    \renewcommand{\arraystretch}{1.1}
    \begin{tabular}{*{7}{c}}
    \br
    Formulation & $1-R^{2}$ &  $n_{\text{DoFs}}$ & $n_{\text{linsys}}$ & $t_{\text{a}}$\,(min) & $t_{\text{s}}$\,(min) & $t_{\text{tot}}$\,(min) \\
    \hline 2D $h$-$\phi$ ref. & -- & \SI{15}{}k & 1108 & \SI{9.3}{} & \SI{2.5}{} & \SI{12}{} \\
    2D $j$-$a$ FCM & \SI{2.3e-4}{} & \SI{4.1}{}k & 994 & \SI{0.4}{} & \SI{0.2}{} & \SI{0.6}{} \\
    2D $h$-$\phi$ FCM & \SI{1.9e-4}{} & \SI{2.7}{}k & 1700 & \SI{0.6}{} & \SI{0.2}{} & \SI{0.8}{} \\
    2D $h$-(full) FCM & \SI{1.4e-4}{} & \SI{6.3}{}k & 989 & \SI{0.5}{} & \SI{0.5}{} & \SI{1.0}{} \\
    3D $t$-$\omega$ FCM & \SI{1.9e-4}{} & \SI{94}{}k & 1019 & \SI{10}{} & \SI{81}{} & \SI{91}{} \\
    3D $h$-$\phi$ FCM & \SI{1.9e-4}{} & \SI{120}{}k & 1021 & \SI{14}{} & \SI{130}{} & \SI{140}{} \\
    3D $h$-(full) FCM & \SI{2.0e-4}{} & \SI{400}{}k & 1023 & \SI{13}{} & \SI{600}{} & \SI{610}{} \\
    3D $h$-$\phi$ res. & \SI{2.5e-3}{} & \SI{630}{}k & 1921 & \SI{59}{} & \SI{3400}{} & \SI{3600}{} \\
    \br
    \end{tabular}
    \label{tab:pc_times}
\end{table*}

Table~\ref{tab:pc_times} additionally gathers details of the numerical performance of the different models. The total computation time, $t_\text{tot}$, the time needed to assemble the finite element matrices, $t_{\text{a}}$, the time needed to solve the linear systems, $t_{\text{s}}$, the number of linear systems solved, $n_{\text{linsys}}$, and $n_{\text{DoFs}}$ are reported. The different homogenized models use the same mesh in 2D and 3D geometries. As expected, the 2D axisymmetric models run significantly faster and require a small number of DoFs. The deviations in performance are larger between the 3D models. The $t$-$\omega$ and $h$-$\phi$ FCMs turn out to be the most efficient in terms of computation time and DoFs. This is explained by the fact that both models simultaneously use homogenization and express the magnetic field in terms of $\phi$ in non-conducting regions. Out of these two formulations, $t$-$\omega$ has the best performance due to its further reduction of DoFs.

From all of the examined models, the 3D resolved model is the least accurate in terms of the $R^{2}$ coefficient and has the highest $n_{\text{DoFs}}$ and $t_{\text{tot}}$. Partly, the high computational cost is due to modelling the entire pancake geometry. The larger error in losses, also shown in the inset of figure~\ref{fig:pc_losses}, is assumed to be due to the mesh used. Mesh refinement for the 3D resolved model up to the results of table~\ref{tab:pc_times} achieves increasing accuracy but further refinement is not considered due to the already significant computational cost.

While it is possible to get shorter computation times, the focus is here set on the verification of the proposed model with results from \cite{Paakkunainen_2025aa} and \cite{Denis_2025aa}. It is noted that the 3D mesh could be made coarser if the mean losses $P$, defined as
\begin{equation}
    P = \frac{2}{T} \displaystyle\int_{T/2}^{T}\, p \,dt,
\end{equation}
is the quantity of interest. For example, with the 3D $t$-$\omega$ FCM, if a relative error of \SI{1.3}{\percent} in the mean losses is acceptable, the number of DoFs could be reduced to \SI{9.6}{}k leading to a total simulation time of \SI{2.1}{\minute}. The relative error is defined as
\begin{equation}
    \relerrpavg = \frac{|P - P_{\text{ref}}|}{P_{\text{ref}}},
    \label{eq:relerr_pavg}
\end{equation}
where $P_{\text{ref}}$ is the mean losses of the reference model. The FCM allows for simpler geometry generation and meshing compared to the resolved models, as the mesh and $\Phi(\alpha)$ can be independent of the CCs in the coil. Also, the cohomology cut generation is simpler for the homogenized models as only one cut is needed per coil whereas, e.g., the 2D reference model requires one cut for each of the conductors. The easier definition of the problem is an advantage of the homogenized models, although not directly visible in the runtimes.

To further demonstrate the computational benefits of the FCM, pancake coils with different numbers of turns are examined in 3D. Table~\ref{tab:pc_scaling} gathers the numerical performance when increasing the number of turns in the coil. For brevity, only the $t$-$\omega$ FCM is considered, and the errors $\relerrpavg$ are again reported with respect to the finely discretized 2D axisymmetric reference model. Coarser meshes are used to speed up the simulations as can be seen when comparing the results with $N_{\text{c}}=20$ to table~\ref{tab:pc_times}, particularly $n_{\text{DoFs}}$ is smaller and the discrepancy in the losses is higher. When denoting the number of structured elements in the local $\alpha$-$\beta$-$\gamma$ coordinates (see figure~\ref{fig:fcm-principle}) as $N_{\alpha} \times N_{\beta} \times N_{\gamma}\,$, the 3D resolved model is meshed with $2 N_{\text{c}} \times 48 \times 60$ elements and the FCM with $\{5,8\} \times 31 \times 50$ elements. The homogenized model is discretized with $N_{\alpha}=5$ elements when $N_{\text{c}}=20$, otherwise $N_{\alpha}=8$. The FCM enables to reduce $N_{\alpha}$ significantly. It is observed that the computational cost of the 3D resolved model increases considerably with the number of turns, whereas the homogenized models show only a mild increase in the number of DoFs and total simulation time. While some advantage is inherent to the symmetries considered for the homogenized model, the 3D $t$-$\omega$ FCM is consistently at least one order of magnitude faster than the 3D resolved model even though the resolved model uses a comparatively coarse mesh. Additionally, the relative error is consistently smaller with the homogenized models. This highlights the advantage of the FCM for simulating coils with a large number of turns.

\begin{table}
    \caption{Computation cost when varying the number of turns in the pancake coil in 3D.}
    \centering
    \renewcommand{\arraystretch}{1.2}
    \begin{tabular}{*{5}{c}}
    \br 
    $N_{\text{c}}$ & Formulation & $n_{\text{DoFs}}$ & $t_{\text{tot}}$\,(min) & $\relerrpavg$ \\
    \hline
    \multirow{2}{*}{$20$} & 3D $t$-$\omega$ FCM & \SI{51}{}k & \SI{20}{} & \SI{1.6}{\percent} \\
     & 3D $h$-$\phi$ res. & \SI{320}{}k & \SI{640}{} & \SI{10}{\percent} \\
    \hline
    \multirow{2}{*}{$30$} & 3D $t$-$\omega$ FCM & \SI{63}{}k & \SI{32}{} & \SI{1.2}{\percent} \\
     & 3D $h$-$\phi$ res. & \SI{400}{}k & \SI{2300}{} & \SI{8.6}{\percent} \\
    \hline
    \multirow{2}{*}{$50$} & 3D $t$-$\omega$ FCM & \SI{64}{}k & \SI{37}{} & \SI{1.0}{\percent} \\
     & 3D $h$-$\phi$ res. & \SI{560}{}k & \SI{3700}{} & \SI{4.9}{\percent} \\
    \hline
    \multirow{2}{*}{$100$} & 3D $t$-$\omega$ FCM & \SI{60}{}k & \SI{39}{} & \SI{1.1}{\percent}\\
     & 3D $h$-$\phi$ res. & - & - & - \\
    \br
    \end{tabular}
    \label{tab:pc_scaling}
\end{table}

\subsection{Stack of Racetrack Coils}

\begin{figure}
    \centering
    \includegraphics[width=0.4\textwidth,trim=0em 2em 0.8em -0.5em,clip]{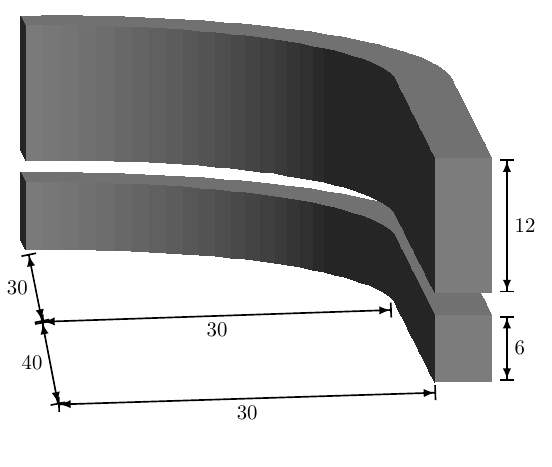}
    \caption{Stack of racetrack coils geometry. The units are in \SI{}{\milli\meter}. The gap between the coils is \SI{2}{\milli\meter}.}
    \label{fig:rt_geom}
\end{figure}

To demonstrate the applicability of the proposed models to general 3D geometries, a stack of 3 racetrack coils with $N_{\text{c}}=50$ turns each is examined. The modelling domain is 1/8 of the stack, as shown in figure~\ref{fig:rt_geom} alongside the dimensions of the coils. A sinusoidal current with the amplitude of \SI{90}{\ampere} at \SI{50}{\hertz} is applied. Again, the continuous voltage function $\Phi(\alpha)$ is approximated with a third-order global polynomial.

For this simulation example, a material fit for $\jc$ of a commercial Fujikura CC from the STEAM Material Library \cite{Bortot_2018ab, steamMaterialLibrary} is used, which takes into account the angle and magnitude of the magnetic field as well as the temperature. For the simulation considering only electromagnetic effects, a constant temperature of $\SI{77}{\kelvin}$ is assumed, and, for reference, $\jc$ at this temperature at \SI{0.5}{\tesla} with field lines parallel to the wide surface of the CC is \SI{6.9e9}{\ampere\per\square\meter}. Using these material fits requires a specially compiled GetDP version CERNGetDP \cite{cerngetdp}.

\begin{figure}
    \centering
    \includegraphics[width=0.5\textwidth,trim=1.2em 0.2em -1.2em 0em,clip]{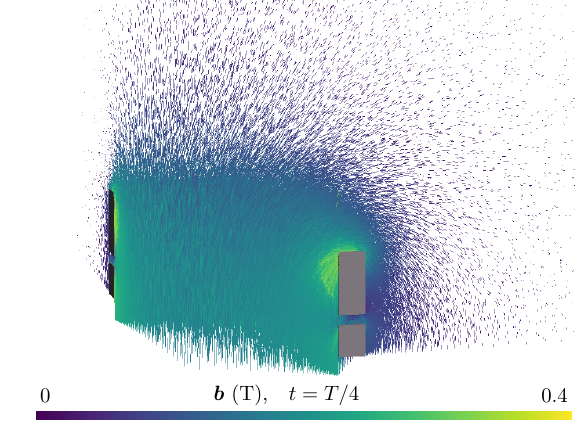}
    \caption{The magnetic flux density distribution for the stack of racetrack coils model at the maximum of the imposed current, computed with the $t$-$\omega$ FCM.}
    \label{fig:rt_b}
\end{figure}

Figure~\ref{fig:rt_b} shows the distribution of the magnetic flux density $\b$ at the maximum of the imposed current for the $t$-$\omega$ FCM. Table~\ref{tab:rt_times} gathers details of the numerical performance of the different FCMs and the reference model in 3D. The reference model considers all the distinct turns of the racetracks while neglecting their spiral structure, consequently assuming concentric turns. This allows the reference model to consider only 1/8 of the geometry. All the homogenized formulations use the same mesh. A good agreement of the mean losses is observed, and the numerical performance is similar as in the pancake coil example. Again, the $t$-$\omega$ FCM has the best performance allowing for the 3D simulation of a stack of racetrack coils in \SI{68}{\minute}. With respect to the reference model, the $t$-$\omega$ FCM provides a speedup factor of 22 and reduces the DoFs by \SI{78}{\percent}.

\begin{table}
    \caption{Computation cost of different formulations for the stack of racetrack coil example.}
    \centering
    \renewcommand{\arraystretch}{1.1}
    \begin{tabular}{*{4}{c}}
    \br
    Formulation & $P$\,(W) & $n_{\text{DoFs}}$ & $t_{\text{tot}}$\,(min) \\
    \hline 3D $t$-$\omega$ FCM & \SI{10.76}{} & \SI{86}{}k & \SI{68}{} \\
    3D $h$-$\phi$ FCM & \SI{10.77}{} & \SI{110}{}k & \SI{85}{} \\
    3D $h$-(full) FCM & \SI{10.79}{} & \SI{330}{}k & \SI{370}{} \\
    3D $h$-$\phi$ ref. & \SI{10.85}{} & \SI{390}{}k & \SI{1500}{} \\
    \br 
    \end{tabular}
    \label{tab:rt_times}
\end{table}

\begin{figure}
    \centering
    \includegraphics[width=0.53\textwidth,trim=0em 0em -3em 0em,clip]{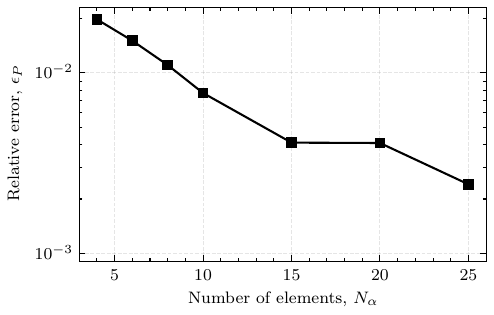}
    \caption{The relative error of the mean AC losses $\relerrpavg$ as a function of the number of elements in the coils in the $\alpha$-direction $N_{\alpha}$ for the stack of racetrack coils example. The results are computed with the $t$-$\omega$ FCM.}
    \label{fig:rt_mesh_refinement}
\end{figure}

We conclude the stack of racetrack coils simulation example by carrying out a mesh refinement study. Figure~\ref{fig:rt_mesh_refinement} shows $\relerrpavg$ as a function of $N_{\alpha}$. The relative errors for the mean AC losses are now computed with respect to the 3D $h$-$\phi$ reference model. The mesh refinement is quantified with $N_{\alpha}$, i.e., the number of equidistant hexahedral elements in the $\alpha$-direction of the local coordinate system, even though the coil mesh is refined in all directions. A reduction in the error of the $t$-$\omega$ FCM is observed with mesh refinement. Already with small $N_{\alpha}$, good accuracies for the mean losses are obtained and, e.g., the results in table~\ref{tab:rt_times} are computed with $N_{\alpha}=10$. If a lower resolution of the field quantities is sufficient, a coarse mesh with $N_{\alpha}=4$ can be used, resulting in $\relerrpavg=\SI{2.0}{\percent}$ with \SI{12}{}k DoFs and a total simulation time of \SI{6.3}{\minute}. 

\section{Conclusions}\label{sec:conclusions}
In this paper, the FCM with magnetic field conforming formulations, i.e., the $h$-(full), $h$-$\phi$, and $t$-$\omega$ formulations, is applied to the simulation of insulated HTS coils. The presented homogenization approach is verified in 2D axisymmetric and 3D geometries through a pancake coil example with comparison to reference models. Additionally, a stack of 3 racetrack coils is examined in 3D to demonstrate the applicability of the FCM to large scale models. The results demonstrate the accuracy and computational efficiency of the FCM by yielding a speedup factor of up to 22 and a reduction of up to \SI{78}{\percent} in DoFs for the stack of racetrack coils simulation. The examined pancake and racetrack coil examples form the basis for studying more complex systems containing insulated HTS coils in both 2D and~3D.

\section*{Acknowledgment}
The work of Elias Paakkunainen is supported by the Graduate School CE in Computational Engineering at the Technical University of Darmstadt. The work of Louis Denis is supported by the F.R.S.-FNRS.

The authors would like to thank Erik Schnaubelt from CERN for his support on the usage of FiQuS Pancake3D module, CERNGetDP, and the STEAM material library.


\section*{References}
\begin{thebibliography}{10}
\providecommand{\url}[1]{#1}
\providecommand{\doi}[1]{\href{https://doi.org/#1}{doi: #1}}
\csname url@samestyle\endcsname
\providecommand{\newblock}{\relax}
\providecommand{\bibinfo}[2]{#2}
\providecommand{\BIBentrySTDinterwordspacing}{\spaceskip=0pt\relax}
\providecommand{\BIBentryALTinterwordstretchfactor}{4}
\providecommand{\BIBentryALTinterwordspacing}{\spaceskip=\fontdimen2\font plus
\BIBentryALTinterwordstretchfactor\fontdimen3\font minus
  \fontdimen4\font\relax}
\providecommand{\BIBdecl}{\relax}
\BIBdecl

\bibitem{Coombs_2024aa}
T.~A. Coombs et al., ``High-temperature superconductors and their large-scale
  applications,'' \emph{Nature Reviews Electrical Engineering}, vol.~1, pp.
  788--801, Dec. 2024. \doi{10.1038/s44287-024-00112-y}.

\bibitem{Bottura_2022ab}
L.~Bottura, S.~Prestemon, L.~Rossi, and A.~V. Zlobin, ``Superconducting magnets
  and technologies for future colliders,'' \emph{Frontiers in Physics},
  vol.~10, Oct. 2022. \doi{10.3389/fphy.2022.935196}.

\bibitem{Hartwig_2024aa}
Z.~S. Hartwig et al., ``The {SPARC} toroidal field model coil program,''
  \emph{{IEEE} Transactions on Applied Superconductivity}, vol.~34, no.~2, pp.
  1--16, Mar. 2024. \doi{10.1109/TASC.2023.3332613}.

\bibitem{Haran_2017aa}
K.~S. Haran et al., ``High power density superconducting rotating machines --
  development status and technology roadmap,'' \emph{Superconductor Science
  and Technology}, vol.~30, no.~12, p. 123002, Dec. 2017. \doi{10.1088/1361-6668/aa833e}.

\bibitem{Sotelo_2022aa}
G.~Goncalves~Sotelo et al., ``A review of superconducting fault current
  limiters compared with other proven technologies,'' \emph{Superconductivity},
  vol.~3, p. 100018, 2022. \doi{10.1016/j.supcon.2022.100018}.

\bibitem{Noe_2026aa}
M.~Noe et al., ``Superconducting high-power cables and lines--development
  status and technology roadmap,'' \emph{Superconductor Science and
  Technology}, vol.~39, no.~2, p. 023501, Feb. 2026. \doi{10.1088/1361-6668/ae15c2}.

\bibitem{Dular_2020aa}
J.~Dular, C.~Geuzaine, and B.~Vanderheyden, ``Finite-element formulations for
  systems with high-temperature superconductors,'' \emph{{IEEE} Transactions on
  Applied Superconductivity}, vol.~30, no.~3, pp. 1--13, Apr. 2020. \doi{10.1109/TASC.2019.2935429}.

\bibitem{Dadhich_2024aa}
A.~Dadhich et al., ``Electromagnetic-thermal modeling of high-temperature
  superconducting coils with homogenized method and different formulations: a
  benchmark,'' \emph{Superconductor Science and Technology}, vol.~37, no.~12,
  p. 125006, Nov. 2024. \doi{10.1088/1361-6668/ad8315}.

\bibitem{Schnaubelt_2023ab}
E.~Schnaubelt, M.~Wozniak, S.~Sch\"{o}ps, and A.~Verweij, ``Electromagnetic
  simulation of no-insulation coils using h\textminus{}$\phi$ thin shell
  approximation,'' \emph{{IEEE} Transactions on Applied Superconductivity},
  vol.~33, no.~5, pp. 1--6, Aug. 2023. \doi{10.1109/TASC.2023.3258905}.

\bibitem{Wang_2025ab}
L.~Wang and J.~Zheng, ``{3D} general simultaneous multi-scale homogeneous
  {T}-{A} model for electromagnetic simulations of large-scale {REBCO} coils
  with complex geometries,'' \emph{Superconductor Science and Technology},
  vol.~38, no.~8, p. 085005, Aug. 2025. \doi{10.1088/1361-6668/adef88}.

\bibitem{Denis_2026aa}
L.~Denis, B.~Vanderheyden, and C.~Geuzaine, ``Simultaneous multi-scale
  homogeneous {H}-{P}hi thin-shell model for efficient simulations of stacked
  {HTS} coils,'' \emph{{IEEE} Transactions on Applied Superconductivity},
  vol.~36, no.~5, pp. 1--7, Aug. 2026. \doi{10.1109/TASC.2026.3652981}.

\bibitem{Dular_2026aa}
J.~Dular, A.~Glock, A.~Verweij, and M.~Wozniak, ``Distributed inter-strand
  coupling current model for finite element simulations of {R}utherford
  cables,'' \emph{Superconductor Science and Technology}, vol.~39, no.~3, p.
  035006, Mar. 2026. \doi{10.1088/1361-6668/ae4a51}.

\bibitem{Basei_2026aa}
R.~Basei, F.~Pase, F.~Lucchini, R.~Torchio, and F.~Toso, ``A structured neural
  {ODE} approach for real-time evaluation of {AC} losses in {3-D}
  superconducting tapes,'' \emph{{IEEE} Transactions on Applied
  Superconductivity}, vol.~36, no.~7, pp. 1--14, Oct. 2026. \doi{10.1109/TASC.2026.3657176}.

\bibitem{Paakkunainen_2024aa}
E.~Paakkunainen, J.~Bundschuh, I.~C. Garcia, H.~De~Gersem, and S.~Sch\"{o}ps,
  ``A stabilized circuit-consistent foil conductor model,'' \emph{{IEEE}
  Access}, vol.~12, pp. 1408--1417, 2024. \doi{10.1109/ACCESS.2023.3346677}.

\bibitem{Zermeno_2013aa}
V.~M.~R. Zermeno, A.~B. Abrahamsen, N.~Mijatovic, B.~B. Jensen, and M.~P.
  S\o{}rensen, ``{Calculation of alternating current losses in stacks and coils
  made of second generation high temperature superconducting tapes for large
  scale applications},'' \emph{Journal of Applied Physics}, vol. 114, no.~17,
  p. 173901, Nov. 2013. \doi{10.1063/1.4827375}.

\bibitem{Berrospe-Juarez_2019aa}
E.~Berrospe-Juarez, V.~M.~R. Zerme\~{n}o, F.~Trillaud, and F.~Grilli,
  ``Real-time simulation of large-scale {HTS} systems: {m}ulti-scale and
  homogeneous models using the {T--A} formulation,'' \emph{Superconductor
  Science and Technology}, vol.~32, no.~6, p. 065003, Apr. 2019. \doi{10.1088/1361-6668/ab0d66}.

\bibitem{Wang_2023aa}
S.~Wang, H.~Yong, and Y.~Zhou, ``Numerical calculations of high temperature
  superconductors with the {J}-{A} formulation,'' \emph{Superconductor Science
  and Technology}, vol.~36, no.~11, p. 115020, Sep. 2023. \doi{10.1088/1361-6668/acfbbe}.

\bibitem{Dos-Santos_2026aa}
G.~dos Santos, B.~M.~O. Santos, F.~J.~M. Dias, N.~Riva, G.~De~Marzi, and
  F.~Grilli, ``{J-A-$\phi$} formulation with homogenizing technique used to
  efficiently model {HTS} cable-in-conduit conductors,'' \emph{{IEEE}
  Transactions on Applied Superconductivity}, vol.~36, no.~3, pp. 1--4, May
  2026. \doi{10.1109/TASC.2025.3638311}.

\bibitem{Zermeno_2014aa}
V.~M.~R. Zerme\~{n}o and F.~Grilli, ``{3D} modeling and simulation of {2G}
  {HTS} stacks and coils,'' \emph{Superconductor Science and Technology},
  vol.~27, no.~4, Apr. 2014. \doi{10.1088/0953-2048/27/4/044025}.

\bibitem{Vargas-Llanos_2022aa}
C.~R. Vargas-Llanos, F.~Huber, N.~Riva, M.~Zhang, and F.~Grilli, ``{3D}
  homogenization of the {T}-{A} formulation for the analysis of coils with
  complex geometries,'' \emph{Superconductor Science and Technology}, vol.~35,
  no.~12, p. 124001, Oct. 2022. \doi{10.1088/1361-6668/ac9932}.

\bibitem{Mataira_2020aa}
R.~Mataira, M.~D. Ainslie, R.~Badcock, and C.~W. Bumby, ``Finite-element
  modelling of no-insulation {HTS} coils using rotated anisotropic
  resistivity,'' \emph{Superconductor Science and Technology}, vol.~33, no.~8,
  p. 08LT01, Jun. 2020. \doi{10.1088/1361-6668/ab9688}.

\bibitem{Paakkunainen_2025aa}
E.~Paakkunainen, L.~Denis, C.~Geuzaine, P.~Rasilo, and S.~Sch\"{o}ps, ``Foil
  conductor model for efficient simulation of {HTS} coils in large scale
  applications,'' \emph{{IEEE} Transactions on Applied Superconductivity},
  vol.~35, no.~5, Aug. 2025. \doi{10.1109/TASC.2024.3517573}.

\bibitem{De-Gersem_2001aa}
H.~De~Gersem and K.~Hameyer, ``A finite element model for foil winding
  simulation,'' \emph{{IEEE} Transactions on Magnetics}, vol.~37, no.~5, pp.
  3427--3432, Sep. 2001. \doi{10.1109/20.952629}.

\bibitem{Dular_2002aa}
P.~Dular and C.~Geuzaine, ``Spatially dependent global quantities associated
  with 2-{D} and 3-{D} magnetic vector potential formulations for foil winding
  modeling,'' \emph{{IEEE} Transactions on Magnetics}, vol.~38, no.~2, pp.
  633--636, Mar. 2002. \doi{10.1109/20.996165}.

\bibitem{Denis_2025aa}
L.~Denis, E.~Paakkunainen, P.~Rasilo, S.~Sch\"{o}ps, B.~Vanderheyden, and
  C.~Geuzaine, ``Magnetic field conforming formulations for foil windings,''
  \emph{{IEEE} Transactions on Magnetics}, vol.~61, no.~8, p. 8400607, Aug.
  2025. \doi{10.1109/TMAG.2025.3584111}.

\bibitem{Dular_1999aa}
P.~Dular, F.~Henrotte, and W.~Legros, ``A general and natural method to define
  circuit relations associated with magnetic vector potential formulations,''
  \emph{{IEEE} Transactions on Magnetics}, vol.~35, no.~3, pp. 1630--1633, May
  1999. \doi{10.1109/20.767310}.

\bibitem{Pellikka_2013aa}
M.~Pellikka, S.~Suuriniemi, L.~Kettunen, and C.~Geuzaine, ``Homology and
  cohomology computation in finite element modeling,'' \emph{{SIAM} Journal on
  Scientific Computing}, vol.~35, no.~5, pp. B1195--B1214, 2013. \doi{10.1137/130906556}.

\bibitem{Rhyner_1993aa}
J.~Rhyner, ``Magnetic properties and {AC}-losses of superconductors with power
  law current---voltage characteristics,'' \emph{Physica {C} --
  Superconductivity and its Applications}, vol. 212, no. 3-4, pp. 292--300,
  1993. \doi{10.1016/0921-4534(93)90592-E}.

\bibitem{Bortot_2020aa}
L.~Bortot et al., ``A coupled {A}-{H} formulation for magneto-thermal
  transients in high-temperature superconducting magnets,'' \emph{{IEEE}
  Transactions on Applied Superconductivity}, vol.~30, no.~5, Aug. 2020. \doi{10.1109/TASC.2020.2969476}.

\bibitem{Dlotko_2019aa}
P.~D\l{}otko, B.~Kapidani, S.~Pitassi, and R.~Specogna, ``Fake conductivity or
  cohomology: Which to use when solving eddy current problems with
  $h$-formulations?'' \emph{{IEEE} Transactions on Magnetics}, vol.~55, no.~6,
  pp. 1--4, 2019. \doi{10.1109/TMAG.2019.2906099}.

\bibitem{Geuzaine_2009ab}
C.~Geuzaine and J.-F. Remacle, ``Gmsh: A {3-D} finite element mesh generator
  with built-in pre- and post-processing facilities,'' \emph{International
  Journal for Numerical Methods in Engineering}, vol.~79, pp. 1309--1331, 2009. \doi{10.1002/nme.2579}.

\bibitem{Dular_1998ac}
P.~Dular, C.~Geuzaine, F.~Henrotte, and W.~Legros, ``A general environment for
  the treatment of discrete problems and its application to the finite element
  method,'' \emph{{IEEE} Transactions on Magnetics}, vol.~34, no.~5, pp.
  3395--3398, Sep. 1998. \doi{10.1109/20.717799}.

\bibitem{Paakkunainen_2026aa}
E.~Paakkunainen, L.~Denis, B.~Vanderheyden, C.~Geuzaine, P.~Rasilo, and
  S.~Sch\"{o}ps, ``{GetDP} models: {H}omogenization of {HTS} coils with the h,
  h-phi, and t-omega foil conductor model,'' \emph{Zenodo}, 2026. \doi{10.5281/zenodo.19163578}.

\bibitem{Atalay_2024aa}
S.~Atalay et al., ``An open-source {3D} {FE} quench simulation tool for
  no-insulation {HTS} pancake coils,'' \emph{Superconductor Science and
  Technology}, vol.~37, no.~6, p. 065005, May 2024. \doi{10.1088/1361-6668/ad3f83}.

\bibitem{Dular_2003ab}
P.~Dular and C.~Geuzaine, ``Modeling of thin insulating layers with dual {3-D}
  magnetodynamic formulations,'' \emph{{IEEE} Transactions on Magnetics},
  vol.~39, no.~3, pp. 1139--1142, May 2003. \doi{10.1109/TMAG.2003.810387}.

\bibitem{Wozniak_2025aa}
M.~Wozniak et al., ``Influence of critical current defect on operation, quench
  detection and protection of a conduction-cooled pancake {REBCO} coil,''
  \emph{{IEEE} Transactions on Applied Superconductivity}, vol.~35, no.~5, pp.
  1--6, Aug. 2025. \doi{10.1109/TASC.2025.3532246}.

\bibitem{Bortot_2018ab}
L.~Bortot et al., ``{STEAM}: A hierarchical co-simulation framework for
  superconducting accelerator magnet circuits,'' \emph{{IEEE} Transactions on
  Applied Superconductivity}, vol.~28, no.~3, Apr. 2018. \doi{10.1109/TASC.2017.2787665}.

\bibitem{steamMaterialLibrary}
{STEAM Team at CERN}. ``{STEAM} material library,'' Accessed: Mar. 23, 2026. [Online]. Available: \url{https://gitlab.cern.ch/steam/steam-material-library}.

\bibitem{cerngetdp}
{STEAM Team at CERN}. ``{CERNGetDP},'' Accessed: Mar. 23, 2026.  [Online]. Available: \url{https://gitlab.cern.ch/steam/cerngetdp}.

\end{thebibliography}
\end{document}